\documentclass{article}
\usepackage{spconf,amsmath,graphicx}
\usepackage{amsthm}
\usepackage{amsmath,amssymb,amsfonts,times,bm}
\newtheorem{theorem}{Theorem}[section]
\newtheorem{lemma}[theorem]{Lemma}

\newtheorem{dfn}[theorem]{Definition}

\newcommand{\mmax}{\operatornamewithlimits{\max}}
\newcounter{mytempeqncnt}
\def\y{\mathbf{y}}

\def\b{\mathbf{b}}
\def\e{\mathbf{e}}
\def\g{\mathbf{g}}
\def\w{\mathbf{w}}
\def\0{\mathbf{0}}
\def\A{\mathbf{A}}

\def\h{\mathbf{h}}
\def\a{\mathbf{a}}

\def\PPhi{\mathbf{\Phi}}
\def\pphi{\boldsymbol{\phi}}

\title{Deterministic Sequences for Compressive MIMO Channel Estimation}
\name{Peng Zhang$^1$, Lu Gan$^2$, Sumei Sun$^3$ and Cong Ling$^1$}

\address{$^1$Department of Electrical and Electronic Engineering\\Imperial College London, UK, email: \{p.zhang12, c.ling\}@imperial.ac.uk\\
$^2$School of Engineering and Design\\Brunel University, UK, email: lu.gan@brunel.ac.uk\\
$^3$Institute for Infocomm Research\\A$^{*}$STAR, Singapore, email: sunsm@i2r.a-star.edu.sg}
\begin{document}

\maketitle
\begin{abstract}
This paper considers the problem of pilot design for compressive multiple-input multiple-output (MIMO) channel estimation. In particular, we are interested in estimating the channels for multiple transmitters simultaneously when the pilot sequences are shorter than the combined channels. Existing works on this topic demonstrated that tools from compressed sensing theory can yield accurate multichannel estimation provided that each pilot sequence is randomly generated. Here, we propose constructing the pilot sequence for each transmitter from a small set of deterministic sequences. We derive a theoretical lower bound on the length of the pilot sequences that guarantees the multichannel estimation with high probability. Simulation results are provided to demonstrate the performance of the proposed method.
\end{abstract}
\section{Introduction}
In MIMO systems channel estimation, the amount of time required to probe the channels between each transmit-receive pair jointly is significantly less than estimating the channels one-at-a-time. However, the receiver observes a superposition of all the channel responses convolved with the corresponding pilot signals with all the transmitters activated simultaneously. Separating this channel responses into individual channel response requires careful design of the pilot sequences. In \cite{QT09}\cite{YChi11}, Golay/Chu sequences were introduced to serve as the pilot sequences, and the overall channel responses can be separated successfully by solving a least square problem as long as the pilot sequences are long enough. When the pilot sequences are relatively short, the respective linear system of equations becomes underdetermined. The theory of compressed sensing provides the mathematical tools for solving underdetermined system of equations provided the solution is known to be sparse. Based on this theory and the assumption that the channels are sparse, \cite{romberg2010sparse} proved that the multichannel estimation problem can be solved if each transmitter sends a different pilot sequence, and each pilot sequence is randomly generated.

In this paper, we propose a design of the pilot signals with deterministic periodic sequences. Our main contributions include: a cyclic shifting pilot sequence assigning method which enables repetitive usage of one sequence among multiple transmitters; and proving the performance guarantee of the proposed measurement matrix in relation to correlation properties of any set of periodic sequences. We also demonstrate that pilot sequences obtained from certain set of FZC sequences can achieve the optimal theoretical guarantee in terms of average case performance. Some numerical results are included to further illustrate the effectiveness of the proposed method. In Section \ref{sec:model}, we explain our system model. In Section \ref{sec:main}, we briefly review coherence based compressed sensing and periodic sequences, and present the main results of this paper. Numerical results are shown in Section \ref{sec:simulation}, followed by the conclusion in Section \ref{sec:conclusion}.


\emph{Notation:} We use boldface to denote matrices and vectors. For an $M\times{N}$ matrix $\A$, $\A(p,q)$ $(1\leq{p}\leq{M}, 1\leq{p}\leq{N})$ denotes the entry on its $p$-th row and $q$-th column. $\A^H$ denotes the complex-conjugate transpose of $\A$. The column-wise submatrix of $\A$, denoted by $\A_{(i:j)}$ for $i<j$, consists of columns from the $i$-th to the $j$-th column of $\A$.
\section{System Model}
\label{sec:model}
We consider the channel estimation in general MIMO systems, where pilot sequences are transmitted from $t$ sources and observed by $r$ receivers. Since the channel estimation scheme is the same for every receiver, we can assume one receiver in our system without loss of generality. We assume that each transmitter emits a pilot sequence $\pphi_{i} = [\phi_{i}(1),\phi_{i}(2),...,\phi_{i}(M)]^{T}\in\mathbb{C}^M$, where $1\leq{i}\leq{t}$. The channel impulse response (CIR) from the $i$-th transmitter to the receiver is denoted by $\h_i=[h_i(1),h_i(2),...,h_i(L)]^{T}$, where $L$ is the maximum channel length over all CIRs. Throughout the paper, we assume that $M\geq{L}$. If only the $i$-th transmitter is activated, the receiver observes the linear convolution between the pilot sequence $\pphi_{i}$ and the corresponding CIR $\h_{i}$ as following:
\begin{equation}
  \y^0_{i}=\pphi_{i}\ast\h_{i}\triangleq\PPhi^{0}_{i}\h_{i}\label{model}
\end{equation}
where
\begin{equation}\label{matrix}
    \PPhi^{0}_{i}=  \begin{bmatrix}
  \phi_{i}(1) & 0 & \cdots & 0 \\
  \phi_{i}(2) & \phi_{i}(1) & \cdots & 0 \\
  \vdots\\
  \phi_{i}(L) & \phi_{i}(L-1) & \cdots & \phi_{i}(1) \\
  \vdots\\
  \phi_{i}(M) & \phi_{i}(M-1) & \cdots & \phi_{i}(M-L+1) \\
  0 & \phi_{i}(M) & \cdots & \phi_{i}(M-L+2) \\
  \vdots\\
  0 & 0 & \cdots & \phi_{i}(M)
  \end{bmatrix}
\end{equation}
and $\y^0_{i}\in{\mathbb{C}^{M+L-1}}$. As proposed in \cite{romberg2010sparse}, the above linear convolution can be translated to circular convolution by adding the first $L-1$ measurements to the last $L-1$ measurements. Then, the shortened measurement vector is given by
\begin{align}
  \y_{i}&=\begin{bmatrix}
  \phi_{i}(L) & \phi_{i}(L-1) & \cdots & \phi_{i}(1) \\
  \cdot\\
  \phi_{i}(M) & \phi_{i}(M-1) & \cdots & \phi_{i}(M-L+1) \\
  \phi_{i}(1) & \phi_{i}(M) & \cdots & \phi_{i}(M-L+2) \\
  \cdot\\
  \phi_{i}(L-1) & \phi_{i}(L-2) & \cdots & \phi_{i}(M)
  \end{bmatrix}\h_{i} \notag\\
   &=\PPhi_{i}\h_{i}, \label{eqn: PPhi_i}
\end{align}
where $\y_{i}\in\mathbb{C}^M$, and $\PPhi_{i}\in\mathbb{C}^{M\times{L}}$ consists of the first $L$ columns of the $M\times{M}$ circulant matrix constructed by the sequence $\pphi_{i}$ with $[\phi_{L},\cdots,\phi_{1},\phi_{M},\cdots,\phi_{L+1}]$ as the first row. When all $t$ transmitters are considered, the overall observation at the receiver can be expressed as the following
\begin{align}
  \y &=\sum^{t}_{i=1}\y_{i}+\w \nonumber\\
     &=\begin{bmatrix} \PPhi_{1} & \PPhi_{2} & \cdots & \PPhi_{t} \end{bmatrix}\h+\w \notag\\
     &\triangleq\PPhi\h+\w, \label{eqn: PPhi}
\end{align}
where $\y\in\mathbb{C}^{M}$, $\w$ denotes the sampled AWGN noise, $\PPhi\in\mathbb{C}^{M\times{N}}$, $N=tL$, is formed by concatenating the partial circulant matrices $\PPhi_{i}$ row-wise, and $\h\in\mathbb{C}^N$ is the combined CIR, i.e. the unknown $tL$-vector consisting of the $t$ channel responses between each transmitter and the receiver. It should be noted that MIMO-OFDM systems are included in our model, since such circular convolution can be obtained directly if the cyclic-prefix length is greater than the maximum channel length over all CIRs for MIMO-OFDM systems.

In this paper, we assume that vector $\h$ is a generic $K$-sparse signal defined as follows:
\begin{dfn}
The support $\text{supp}(\h)$ of size $K$ is selected uniformly at random in $\{1,2,...,N\}$; and the normalized directions $(\h/|\h|)$ of the nonzero coefficients of $\h$ are independent random variables uniformly distributed on the unit circle.
\end{dfn}
When the length of the pilot sequences $M$ is shorter than $N$, estimating the channel response is equivalent to solving an underdetermined system of equations, which is closely related to recent work in compressed sensing.

\section{Compressed Sensing and Pilot Design}
\label{sec:main}
In this section, we present a novel compressed sensing based pilot design scheme for MIMO channel estimation. We first review the coherence property for compressed sensing. Then, we provide a brief introduction on periodic sequences with good correlation properties. Finally, we propose the measurement matrix construction and prove its theoretical guarantees.

\subsection{Compressed Sensing}
Compressed sensing essentially proved that the signal of interest can be stably reconstructed with far less measurements as long as its sparsity is below some threshold level $K$. In terms of average-case performance, the threshold sparsity $K$ is closely related to the coherence property of the measurement matrix. The coherence-based performance guarantee solely depends on two fundamental parameters of the measurement matrix: coherence and spectral norm.

We assume without loss of generality that the measurement matrix $\PPhi$ has unit-normed columns, and denote by \mbox{$\PPhi(:,i)$} the $i$-th column of $\PPhi$ $(\|\PPhi(:,i)\|_{2}=1)$. Then, its coherence $\mu(\PPhi)$ is the largest absolute inner product between any two columns of $\PPhi$:
\begin{equation}\label{coherence}
    \mu(\PPhi) = \mmax_{1\leq{i\neq{j}}\leq{N}}|\langle\PPhi(:,i),\PPhi(:,j)\rangle|.
\end{equation}
It can be shown that $\mu(\PPhi)\in[\sqrt{\frac{N-M}{M(N-1)}},1]$, where the lower bound is known as the Welch bound.

The spectral norm $\|\PPhi\|_2$ of the matrix $\PPhi$ reflects how much the matrix can expand a unit-norm coefficient vector. A lower bound of the spectral norm of the matrix can be obtained by using H$\ddot{\text{o}}$lder's inequality:
\begin{equation}\label{spectral norm}
    \|\PPhi\|^2_2=\|\PPhi^H\PPhi\|_2\geq{M^{-1}}\text{trace}(\PPhi^H\PPhi)=\frac{N}{M}.
\end{equation}
In the noiseless case (i.e, $\w=\0$), the coherence-based recovery guarantee can be summarized in the following theorem.
\begin{theorem}[\cite{Tropp08on}]\label{thm: BP generic}
Suppose the signal $\h$ is taken from the generic $K$-sparse model. Write $\y=\PPhi\h$ and fix $c_0\geq{1}$. If
\begin{equation}\label{BP generic}
    \sqrt{\mu^2K\cdot{c_0}\log{N}}+\frac{K}{N}\|\PPhi\|^2_2\leq{C},
\end{equation}
where $C>0$ is a constant, then $\h$ is the unique solution to $l_1$-minimization program (basis pursuit) with probability $1-N^{-c_0}$.
\end{theorem}
In the noisy case, \cite{Candes09near} proved that generic $K$-sparse signal $\h$ can be identified using the unconstrained LASSO with very large probability provided that the amplitudes of the nonzero coefficients $h_i$ stand above the noise. In both theorems, it can be observed that smaller coherence $\mu(\PPhi)$ and spectral norm $\|\PPhi\|_2$ of the measurement matrix support better recovery guarantees.

\subsection{Periodic sequences}
Let $X$ denote a set of $T$ sequences of period $M$, then the periodic crosscorrelation function $\theta(\a_u,\a_v)(l)$ with lag $l$ is defined as
\begin{equation}\label{xcorrelation}
    \theta(\a_u,\a_v)(l)=\sum^{M-1}_{k=0} \a_u(k)\a^{*}_{v}(k+l), \quad l\in\mathbb{Z},
\end{equation}
where $\a^{*}_v$ denotes the complex conjugate of $\a_v$, and $\a_v(k)=\a_v(k+M)$ for all $k\in\mathbb{Z}$. The periodic autocorrelation function of the sequences with lag $l$, $\theta(\a_u)(l)$, is just $\theta(\a_u,\a_u)(l)$. The maximum periodic crosscorrelation magnitude $\theta_c$ and the maximum out-of-phase periodic autocorrelation magnitude $\theta_a$ are defined by
\begin{align*}
  \theta_c&=\mmax\{|\theta(\a_u,\a_v)(l)|: \a_u,\a_v\in{X}, u\neq{v}, 0\leq{l}\leq{M-1}\}. \\
  \theta_a&=\mmax\{|\theta(\a_u)(l)|:\a_u\in{X}, 0<{l}\leq{M-1}\}
\end{align*}
Generally, a set of periodic sequences with good correlation properties means both $\theta_c$ and $\theta_a$ are small. However, as Sarwate proved in \cite{DVSarwate79} that the two parameter are constrained by the following
\begin{equation}\label{eqn: sarwat bounds}
    (\frac{\theta_c^2}{M})+\frac{M-1}{M(T-1)}(\frac{\theta_a^2}{M})\geq\frac{1}{M^2},
\end{equation}
which holds for any set $X$ of $T$ sequences of period $M$ satisfying $\theta(\a_u)(0)=1$ for all $\a_u\in{X}$. (Note that this is a normalized version of the result in \cite{DVSarwate79}, and all types of sequences considered in this paper are normalized by $\frac{1}{\sqrt{M}}$ compared with their standard definitions.) Sarwate also proposed a method for constructing a set of Frank-Zadoff-Chu (FZC) sequences that satisfy \eqref{eqn: sarwat bounds} with equality:
\begin{theorem}[\cite{DVSarwate79}]\label{theorem: FZC constrct}
Let $p$ denote the smallest prime divisor of the odd integer $M$, and let $u_i$ denote the multiplicative inverse of $i$ modulo $M$, $1\leq{i}\leq{p-1}$. Then $\theta_c={\frac{1}{\sqrt{M}}}$ and $\theta_a=0$ for the set $X=\{\a_{u_i}|1\leq{i}\leq{p-1}\}$ of FZC sequences defined by
\begin{equation}\label{FZC}
    \a_u(k)=\left\{
  \begin{array}{l l}
    \frac{1}{\sqrt{M}}\exp(j\pi\frac{uk^2}{M}) & \quad \text{if $M$ is even,}\\
    \frac{1}{\sqrt{M}}\exp(j\pi\frac{uk(k+1)}{M}) & \quad \text{if $M$ is odd.}
  \end{array} \right.
\end{equation}
\end{theorem}
Other types of sequences that can consist of a set of sequences with good correlation properties include Gold sequence, Kasami sequence, $m$-sequence, etc\cite{Sarwate80}.
\subsection{Pilot Design}
In this subsection, we propose our pilot design method for general MIMO channel estimation model as given by (\ref{eqn: PPhi}). 
Given the length of pilot sequences $M$, we can obtain a set of $T$ sequences $X$ with $\theta(\a_u)(0)=1$ for all $\a_u\in{X}$. Assume we can find an integer $q$ that satisfies the conditions: $qM=tL$, and $q\leq{T}$, where $L$ is  the maximum channel length, and $t$ is the total number of transmitters. A subset $B$ of $X$ with size $q$, denoted by $B=\{\b_1,\b_2,\cdots,\b_q\}$, consists of the sources for our pilot sequences.

Suppose the $M\times{M}$ circulant matrix $\A$ constructed by $\b_i$ as the first row is defined as $\A^{\b_i}$, and $\b^{+m}_{i}$ represents the sequence obtained by left cyclic shift $\b_i$ by $m$ positions, i.e. $\b^{+m}_{i}=[\b_i(m+1),\b_i(m+2),\cdots,\b_i(M),\b_i(1),\cdots,\b_i(m)]$. Then, we can obtain the following equation
\begin{equation}\label{shift circulant matrix}
    \A^{\b_{i}}_{(m+1:m+n)}=\A^{\b^{+m}_{i}}_{(1:n)}, \quad {\text{if $m+n\leq{M}$}}.
\end{equation}
For any two sequences $\e$ and $\g$ with length $M$, suppose $f(\e)=\g$ is a bijective function that assigns each element of $\e$ to a position in $\g$, such that $[\e(1),\e(2),\cdots,\e(M)]=[\g(L),\cdots,\g(M),\g(1),\cdots,\g(L+1)]$. Then, each transmit pilot sequence $\pphi_i$ $(1\leq{i}\leq{t})$ can be designed as:
\begin{equation}\label{mapping function}
    \pphi_i=f(\b^{+\alpha(i)}_{\beta(i)})
\end{equation}
where $\alpha(i)=(i\bmod{\frac{M}{L}}-1)L$ and $\beta(i)=\lceil{i}\frac{L}{M}\rceil$. For example, the pilot sequence for the first transmitter is constructed by sequence $\b_1$, $[\b_1(1),\b_1(2),\cdots,\b_1(M)]=[\pphi_1(L),\cdots,\pphi_1(M),\pphi_1(1),\cdots,\pphi_1(L+1)]$, and we have $\PPhi_1=\A^{\b_1}_{(1:L)}$, where $\PPhi_1$ is defined by (\ref{eqn: PPhi_i}). In other words, every matrix $\PPhi_i$ can be represented by the following
\begin{equation}\label{equivalent}
    \PPhi^{i}_{(1:L)}=\A^{\b^{+\alpha(i)}_{\beta(i)}}_{(1:L)}.
\end{equation}
The structure of the measurement matrix $\PPhi$ can be formulated as (\ref{PPhi structure}). By mapping the cyclic shifted sequence to certain group of pilot sequences, the resultant measurement can be regarded as row-wise concatenation of $q$ circulant matrices, where each circulant matrix is constructed by a periodic sequence $\b_u\in{B}$.
\begin{figure*}[!t]
\normalsize
\setcounter{mytempeqncnt}{\value{equation}}
\setcounter{equation}{13}
\begin{align}
\PPhi &=\begin{bmatrix} \PPhi_{1} & \PPhi_{2} & \cdots & \PPhi_{t} \end{bmatrix}\notag \\
     &=\begin{bmatrix} \A^{\b_1}_{(1:L)} & \A^{\b^{+L}_1}_{(1:L)} & \A^{\b^{+2L}_1}_{(1:L)} & \cdots & \A^{\b^{+(\frac{M}{L}-1)L}_1}_{(1:L)} & \A^{\b^{+L}_2}_{(1:L)} & \cdots & & \A^{\b^{+(\frac{M}{L}-1)L}_q}_{(1:L)} \end{bmatrix}\notag \\
     &=\begin{bmatrix} \A^{\b_1}_{(1:L)} & \A^{\b_1}_{(L+1:2L)} & \A^{\b_1}_{(2L+1:3L)} & \cdots & \A^{\b_1}_{(M-L+1:M)} & \A^{\b^{+L}_2}_{(1:L)} & \cdots & & \A^{\b_q}_{(M-L+1:M)} \end{bmatrix}\notag \\
     &=\begin{bmatrix} \A^{\b_1} & \A^{\b_2} & \cdots & \A^{\b_q} \end{bmatrix}\label{PPhi structure}
\end{align}
\setcounter{equation}{14}
\hrulefill
\vspace*{4pt}
\end{figure*}

The following Lemma presents the coherence and spectral norm bound of the measurement matrix constructed by the proposed method.
\begin{lemma}\label{my lemma}
Consider a MIMO channel estimation system given by (\ref{eqn: PPhi}). For any $M$ and $q$, suppose the following condition holds: $qM=tL=N$ and $q\leq{T}$. If each pilot sequence is assigned by (\ref{mapping function}), then the coherence and spectral norm of the concatenation matrix $\PPhi$ obeys
\begin{gather}
\sqrt{\frac{N-M}{M(N-1)}}\leq\mu{(\PPhi)}\leq \mmax\{\theta_a,\theta_c\}\\
\frac{N}{M}\leq\|\PPhi\|^2_2\leq\frac{N}{M}(1+\theta_a(M-1)) \label{lemma norm}
\end{gather}
\end{lemma}
The lower bounds for both parameters are from Section 3.1. The upper bound for the coherence can be proved from the observation that the inner product between any two different columns in the matrix is either the out-of-phase periodic autocorrelation of any sequence in set $B$, or the periodic crosscorrelation of two sequences in set $B$. The upper bound for the spectral norm can be derived by using Gershgorin circle theorem\cite{gershgorin1931uber}. The two upper bounds for the measurement matrices constructed by some sets of sequences are summarized in Table $1$.
\begin{table}\label{table}
  \centering
  \caption{Coherence and spectral norm bounds of $\PPhi$ constructed by different sequences ($s\in\mathbb{Z}^{+}$ )}\label{table 1}
\vspace{3mm}
\begin{tabular}{|c|c|c|}
  \hline
  Sequence Type & Coherence & $\|\PPhi\|^2_2$ \\
  \hline
   & &\\[-1.5ex]
  FZC sequence & $\frac{1}{\sqrt{M}}$ & $\frac{N}{M}$ \\[1ex]
  \hline
  Gold sequence & & \\[-1ex]
  $M=2^s-1$ & \raisebox{1.5ex}{$\frac{\sqrt{2(M+1)}}{M}=\gamma$} & \raisebox{1.5ex}{$\frac{N}{M}(1+\gamma(M-1))$} \\
  \hline
  Kasami sequence & & \\[-1ex]
  $M=2^s-1$ & \raisebox{1.5ex}{$\frac{\sqrt{M+1}}{M}=\gamma$} & \raisebox{1.5ex}{$\frac{N}{M}(1+\gamma(M-1))$} \\
  \hline
\end{tabular}
\end{table}
When we choose our pilot sequences from a set of FZC sequences obtained by Theorem 3.2, the spectral norm of the resultant measurement matrix meets the lower bound and the coherence approaches to the Welch bound when $N\gg{M}$. By applying Theorem 3.1, we have:
\begin{theorem}
Suppose the combined CIR $\h$ obeys the generic $K$-sparse model. Given $\y=\PPhi\h$, where $\PPhi$ is the measurement matrix constructed by the proposed method with a set of FZC sequences satisfying $\theta_c=\frac{1}{\sqrt{M}}$, and fix $c_0\geq{1}$. If
\begin{equation}\label{BP generic}
    \sqrt{\frac{K}{M}\cdot{c_0}\log{N}}+\frac{K}{M}\leq{C},
\end{equation}
where $C>0$ is a constant, then $\h$ is the unique solution to $l_1$-minimization program with probability $1-N^{-c_0}$. In other words, the sparsity level that allows for highly accurate channel estimation is $K=\mathcal{O}(M)$.
\end{theorem}

In the noisy case, the theoretical guarantee of the measurement matrix can be obtained by the combination of Theorem \ref{my lemma} and Theorem 1.2 in \cite{Candes09near}. Although our measurement matrix proves weaker theoretical guarantee (coherence) than the one proposed in \cite{romberg2010sparse}\cite{JRomberg09multiple} (isometric property), our construction of measurement matrix is based on deterministic sequences; and the number of different pilot sequences required in our system is reduced by $\frac{M}{L}$ times because of the method of cyclic shifting. Note that similar cyclic shifting method has been proposed in \cite{YChi11}; however, its pilot sequences designed on frequency domain works solely for MIMO-OFDM system, and the corresponding measurement matrix is a well-conditioned column-wise sub matrix of a circulant matrix $(M\geq{N})$, instead of the concatenation matrix in this paper $(M\leq{N})$. \cite{Apple11} proposed estimating the channel in OFDM systems with deterministic pilots, but it is only designed for single transmitter instead of MIMO.

Although the generic $K$-sparse signal model of the combined CIR $\h$ is a reasonable and mild assumption, we can further relax this assumption by incorporating other coherence-based theoretical guarantee \cite{CTLi12}, which also only depends on the coherence and spectral norm. In the case of $M=tL$, our proposed measurement matrix would be a circulant matrix, then random subsampling can be applied to reduced the number of measurements, which can be easily proved by Theorem 3 and Lemma 1 in \cite{LK13}.
\section{Numerical Results}
\label{sec:simulation}
\begin{figure}[htb]
\begin{minipage}[b]{1.0\linewidth}
  \centering
  \centerline{\includegraphics[width=9.0cm]{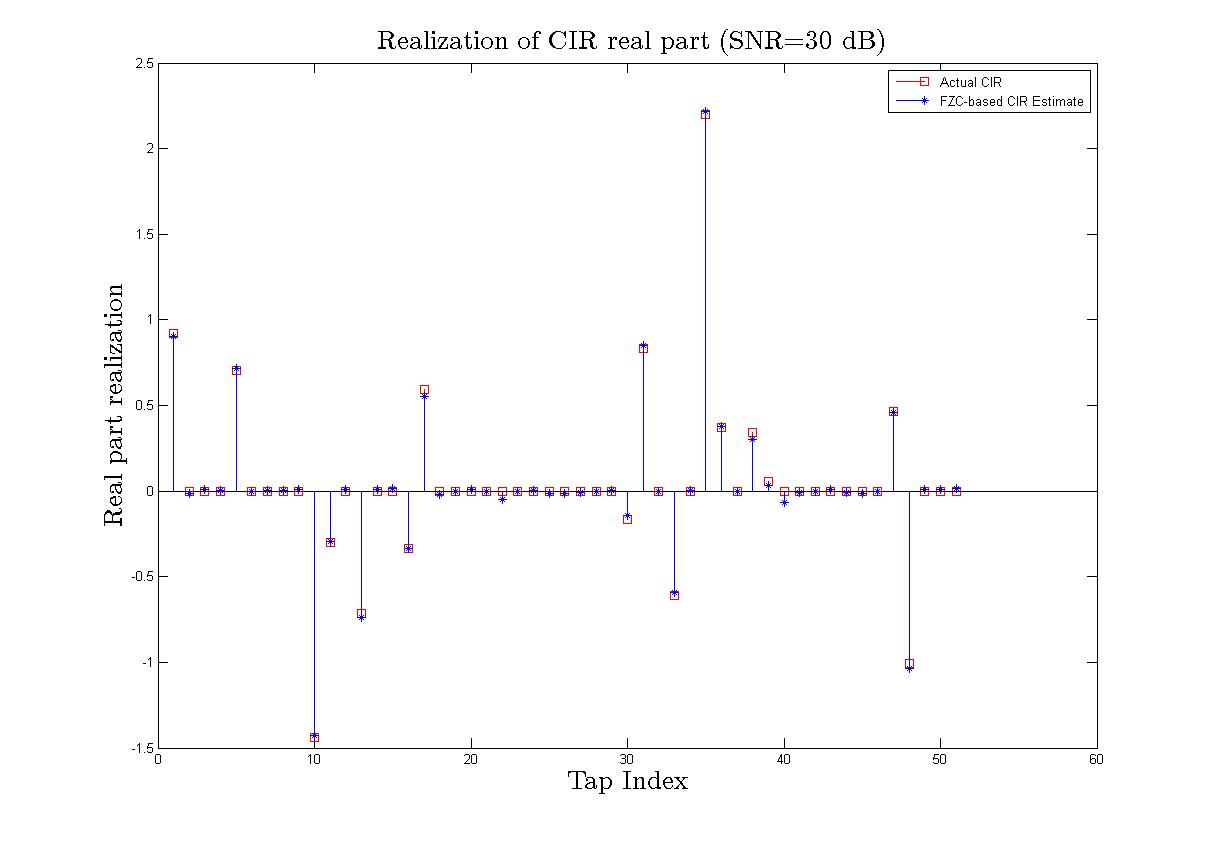}}
\end{minipage}
\caption{Real part of one CIR and its estimates.}
\label{fig:real}
\end{figure}

\begin{figure}[htb]
\begin{minipage}[b]{1.0\linewidth}
  \centering
  \centerline{\includegraphics[width=9.0cm]{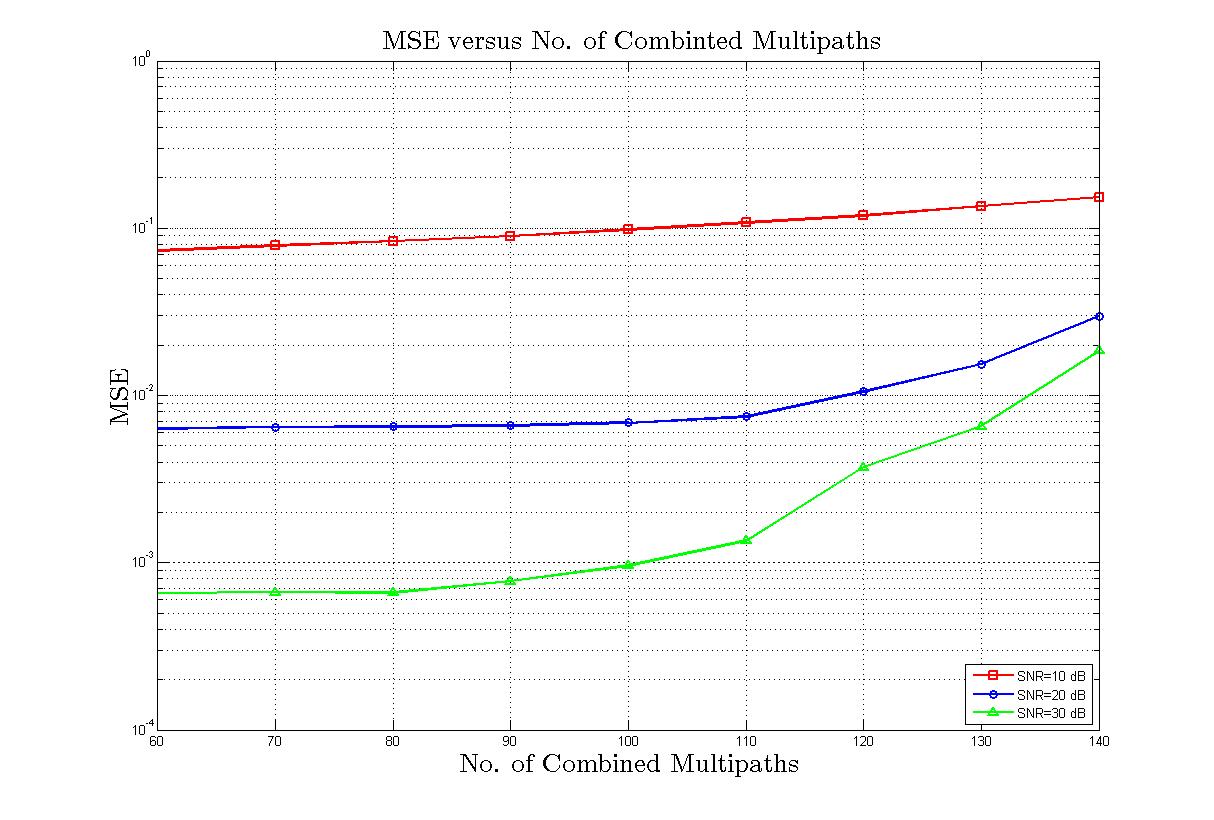}}
\end{minipage}
\caption{MSE performance versus number of combined multipaths.}
\label{fig:mse}
\end{figure}

We have simulated an MIMO system with $M=255$ and $L=51$. The channels between $t=10$ transmitters and one receiver are estimated simultaneously, with measurement SNRs ranging from $10$ dB to $30$ dB. The pilot sequences for each transmitter is assigned by the proposed method. At each SNR, we vary the number of combined multipaths (i.e. total multipaths between the transmitters and the receiver) from $60$ to $140$, which corresponds to the increasing of average number of multipaths per channel from $6$ to $14$. For all simulations, $500$ trials have run using the FPC-AS algorithm \cite{hale07fixed}. Fig.\ref{fig:real} shows the real part of a single CIR realization and its estimates in the presence of SNR$=30$ dB. It shows that the proposed pilot design method successfully detect and estimate the amplitude of a channel with $15$ multipaths. Note that the other $9$ channels is successfully estimated at the same time. Fig. \ref{fig:mse} illustrates the mean squared error ($\text{MSE}={\|\h-\hat{\h}\|_2^2}/{N}$) performance, where $\hat{\h}$ is the estimation on the combined CIR $\h$.
\section{Conclusion}
\label{sec:conclusion}
In this paper, we proposed a new pilot design method based on cyclic shifted periodic sequences. The resultant measurement matrix is proved to satisfy the coherence property for compressed sensing. We derived a general expression for the bounds of coherence and spectral norm in terms of the two correlation parameters for any set of periodic sequences. The optimal theoretical guarantee of the measurement matrix constructed by FZC sequences has been demonstrated by both theoretical analysis and numerical simulations.

\bibliographystyle{IEEEbib}
\bibliography{strings,refs}

\end{document}